\documentclass[runningheads]{llncs}
\usepackage[T1]{fontenc}
\usepackage{graphicx}
\usepackage{booktabs}
\begin{document}
\title{Towards a  Virtual Reality Visualization of Hand-Object Interactions to Support Remote Physical Therapy}
\titlerunning{VR Motion Visualization (VRMoVi) for Remote Physical Therapy}
%
\author{Trudi Di Qi\orcidID{0000-0002-0467-6628} \and
LouAnne Boyd\orcidID{0000-0002-9244-6581} \and
Scott Fitzpatrick \and
Meghna Raswan \and
Franceli L. Cibrian\orcidID{0000-0002-7084-6904} }
\authorrunning{T.D. Qi et al.}
%
\institute{Fowler School of Engineering, Chapman University, Orange CA 92866, USA
\email{\{dqi,lboyd,sfitzpatrick,raswan,cibrian\}@chapman.edu}}

\maketitle              
\begin{abstract}
Improving object manipulation skills through hand-object interaction exercises is crucial for rehabilitation. Despite limited healthcare resources, physical therapists propose remote exercise routines followed up by remote monitoring. However, remote motor skills assessment remains challenging due to the lack of effective motion visualizations. Therefore, exploring innovative ways of visualization is crucial, and virtual reality (VR) has shown the potential to address this limitation. However, it is unclear how VR visualization can represent understandable hand-object interactions. To address this gap, in this paper, we present VRMoVi, a VR visualization system that incorporates multiple levels of 3D visualization layers to depict movements. In a 2-stage study, we showed VRMoVi's potential in representing hand-object interactions, with its visualization outperforming traditional representations, and detailed features improved the hand-object interactions understanding. This study takes the initial step in developing VR visualization of hand-object interaction to support remote physical therapy. 


\end{abstract}
\section{Introduction}
Upper limb rehabilitation typically entails executing specific functional activities like reaching, grasping, moving, and handling objects (i.e., hand-object interaction) \cite{ploderer_2016}. As a key motor skill in physical therapy, improving object manipulation skills is crucial to enhance patients' dexterity and coordination, thus facilitating their daily activities with increased ease and assurance \cite{levin_2015}.

Limited healthcare resources result in a long waiting time and delayed feedback for in-person physical therapy. To address this, physical therapists suggest patients perform remote exercise routines followed by phone, video calls, or even chat \cite{winters_2007}. Unfortunately, with this communication, therapists can only view one angle of the movements, based on the patient’s camera position, which may result in inaccurate assessments \cite{moya_2011}. Therefore, 3-dimensional (3D) visualizations are needed to effectively present hand-object interactions to improve the motion skill assessment. 

Virtual reality (VR) technologies have emerged as promising tools for remote physical therapy showing promising results in improving balance, gait, and motor function in patients with neurological conditions \cite{levin_2015,postolache_2021,jun_2023}. Additionally, recent works take advantage of immersive analytics using VR \cite{kloiber_2020,bschel_2021} to enhance understanding of complex human motion data. However, it is unclear what visualization methods should represent hand motion and object interaction information. Our research aims to develop and evaluate a VR-based visualization method with increasing level-of-detail 3D visualization features to depict hand-object interactions. Our \textbf{contributions} include: 
\begin{itemize}
    \item The VRMoVi prototype, a VR visualization system incorporating multiple levels of 3D visualization layers to depict motion data.
    \item Empirical evidence of a two-stage study with 24 experts in visualizations showing that VRMoVi outperformed the traditional 2D display-based visualization, and adding fine-grained motion features, such as hand positions and rotations, improved their understanding of the data.
\end{itemize}

In the next section, we explore current visualization methods for human limb movement and highlight the importance of enhancing understanding of effective 3D visualizations for remote therapy.

\section{Related Work}
\subsection {Remote Monitoring Strategies in Physical Therapy}  
Remote physical therapy (i.e., telerehabilitation) has gained attention, leveraging advancements in videoconferencing, wearable sensors, and VR/augmented reality (AR), offering promising results for clinical practice and research \cite{jun_2023}. Telerehabilitation uses videoconferencing platforms to support real-time, high-definition video and audio patient and therapist interaction, enabling remote assessment and feedback \cite{winters_2007}. However, they only have a 2D perspective, restricting therapists' ability to assess 3D spatial information of movements \cite{moya_2011}. To overcome this limitation,  depth cameras have been used to get a 3D interpretation of  movements and postures \cite{clark_2012}, but they may still be affected by lighting, positioning, and occlusion \cite{obdrzlek_2012}

Another approach to support remote physical therapy is using wearable devices incorporated into smartphones or wristbands. Those devices can be worn on different body parts to track movements \cite{patel_2012} speed and joint angular motion, providing continuous and objective measures \cite{ploderer_2016a}. Once processed and interpreted, these insights shape personalized therapy plans \cite{dobkin_2011}. 

Recently, VR/AR have emerged as promising tools for remote physical therapy allowing manipulation and control of the therapy environment \cite{jun_2023} (e.g., adjust difficulty levels and provide real-time feedback \cite{postolache_2021}). VR/AR provides interactive and enjoyable experiences \cite{levin_2015} and shows promising results in improving balance, gait, and motor function in patients with neurological conditions \cite{postolache_2021}. This body of work has shown that these technologies can effectively support remote physical therapy; however, it is unclear what type of VR visualization can better represent hand-object interactions. 

\subsection{Motion Data Visualization Approaches}
To facilitate evaluating patients' condition and progress remotely, it is essential to provide physical therapists with comprehensive and interpretable visualizations of patients' movements \cite{ploderer_2016}. Traditional movement visualization includes 2D scatter plots dashboards \cite{ploderer_2016,ploderer_2016a,rawashdeh_2022} mainly of wearable sensors data, supplemented with information such as movement length, velocity, and joint angles. However, this visualization lacks 3D movement information, making it challenging to evaluate the patient’s condition and progress \cite{moya_2011}. 

In physical therapy, computer animations create a simplified animated human skeleton to mimic the patient’s movements in 3D \cite{rado_presentation_2009,moya_2011,postolache_2021}. This can be useful for examining postures and exercises. For example, squat exercises have been represented with computer animations using spheres for joints and sticks for legs \cite{rado_presentation_2009}. This approach allowed therapists to identify the correct positions of the knees. However, animations lack detailed information about movement trajectory and rotations. On the other hand, 3D trajectories have been used in motion analysis applications \cite{kloiber_2020}, mainly for player behavior analysis in sports \cite{sacha_2017} and computer games \cite{kepplinger_2020}. 

Immersive analysis using VR/AR has become an alternative to traditional human motion analysis, enabling 3D visualization of complex motion data beyond 2D displays \cite{kloiber_2020,bschel_2021,reipschlger_2022}. For example, a VR-based visual analysis system using 3D trajectories and avatar animation has been used to represent large-scale movements (e.g., walking)\cite{kloiber_2020}. However, it is unclear how VR visualizations could be applied to hand-object interactions (e.g., picking up, tossing). These interactions demand detailed insights into hand trajectory, rotation, and object interaction, crucial for understanding upper limb movement rehabilitation \cite{ploderer_2016}. In this work, we fill this gap by proposing novel ways of 3D visualization with levels of detail to show the hand-object interactions.

\begin{figure}
\includegraphics[width=\textwidth]{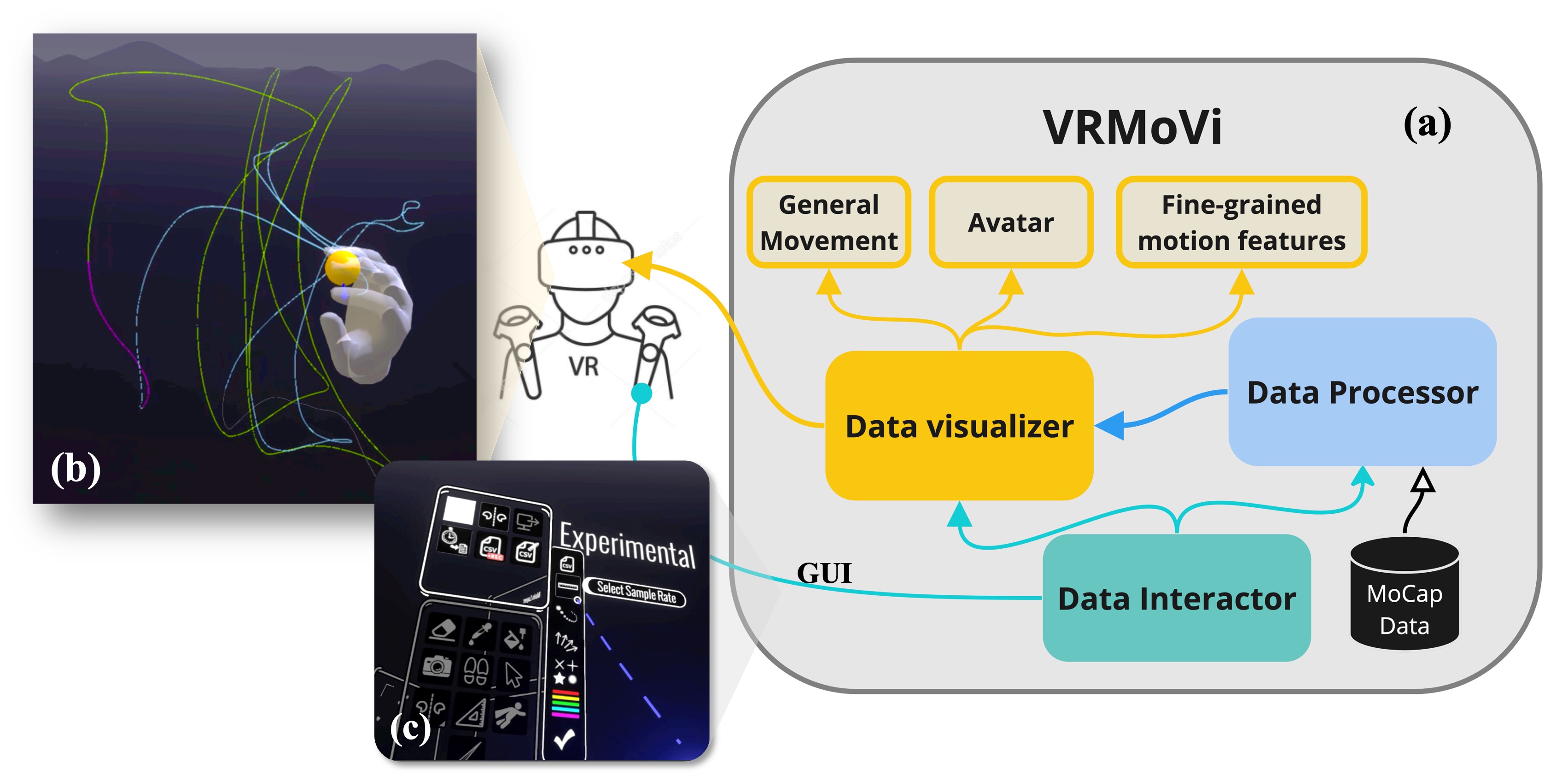}
\caption{(a) VRMoVi consists of three major modules: 1) a data visualizer, 2) a data processor, and 3) a data interactor. (b) illustrates hand motion and object interaction using general movement (GM) and hand-object avatar (A) layers. (c) Shows the GUI (data interactor) allowing the user to interact with the data visualizer and processor.
} 
\label{vrmovi}
\end{figure}

\section{VRMoVi}
This work introduces VRMoVi (Virtual Reality Motion Visualization), a VR-based visualization system with increasing level-of-detail 3D visualization layers: 3D trajectories, avatar animation, and fine-grained visual symbols showing hand positions and rotations to depict hand movements and various hand-object interactions. VRMoVi consists of three  modules to assist the user (e.g., physical therapists) in visualizing human motion data (Fig.\ref{vrmovi}):

\textbf{Data visualizer.} Currently, the data visualizer provides three 3D visualization layers with increasing levels of detail: 
\begin{enumerate}
    \item \textbf{General-Movement (GM)} layer for demonstrating the general movement trajectories of the hand/object (Fig.\ref{vrmovi} (b));
    \item \textbf{Hand-object Avatar (A)} layer for animating the interaction between the 3D models of hands and objects over time  to provide context (Fig.\ref{vrmovi} (b));
    \item \textbf{Fine-grained motion (F)} layer for showing the detailed position and rotation of the hand at each time step to provide specific patterns on a smaller scale (Fig.\ref{visualization} (d))
\end{enumerate}

Visualization layers are developed on top of an open-source VR painting program, OpenBrush \cite{openbrush}, based on Unity \cite{unity}. We use OpenBrush’s painting brushes to realize general movement and fine-grained motion layers. To render the hands’ motion trajectories more clearly and smoothly (see Fig.\ref{vrmovi}(b)), we use translucent 3D curves. The hand position and rotation for each time step are rendered using visual symbols like dots and arrows. The animated hand-object avatar layer is realized by rigid body transformation of 3D models of two hands and objects. Both hands and objects are transformed based on the input hand and object position and rotation data.
 
\textbf{Data interaction.} Users can interact with the VR environment and 3D visualizations through data interactors consisting of hand-held controllers and a graphical user interface (GUI; Fig.\ref{vrmovi} (c)). The system allows users to adjust the density of input motion data or choose any visualization layer(s) to draw the data. 
 
\textbf{Data processor.}  To allow the user to import motion-capture data into VR, we implement a data processor module based on Python data analysis libraries (e.g., Scikit-learn \cite{pedregosa_2011}), allowing users to process the data (e.g., down-sampling) through the data interactor (e.g., dragging a slider to update the resolution of data points) and visualize results in real-time.


    

\section{Evaluation Study}

\subsection{Participants}
We conducted a 2-stage user study with 24 individuals (ages 18-23, 9 female and 15 males) with backgrounds in Computer Science (100\%), Data visualization (16\%), VR (42\%), and Human-Computer Interaction (100\%). In this project stage, we recruited technology design and visualization experts to get initial feedback on the proposed visualization and to build a  robust prototype before conducting sessions with clinicians and experts in physical therapy.


\subsection{Study design}
The study aimed to explore the following research questions (RQ): 
\begin{itemize}
    \item \textbf{RQ1}. \textit{Can participants differentiate hand and object trajectories?}
    \item \textbf{RQ2}. \textit{Can participants interpret the hand-object interaction through visualizations?}
    \item \textbf{RQ3}. \textit{What challenges and opportunities can VR visualization offer for hand-object interaction representations?}
\end{itemize}

We conducted a within-subject quantitative evaluation to respond to RQ1 and RQ2 (\textbf{Stage 1)}; and to address RQ3, we conducted a qualitative study consisting of a brief semi-structured interview of a VR visualization containing various visual symbols spanning from a macro overview to micro details of hand movement and object interactions (\textbf{Stage 2)}. 

\subsubsection{Visualization Methods}
\begin{figure}
\includegraphics[width=\textwidth]{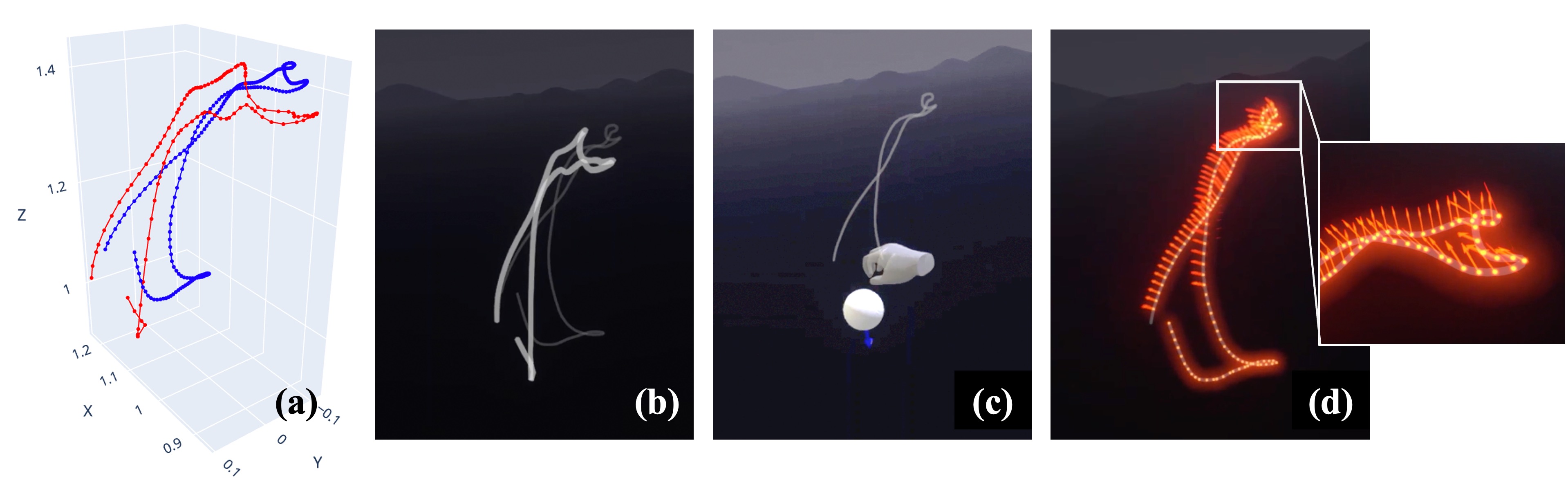}
\caption{Visualizations of hand movement and object interaction for the “Picking-up” scenario showing both hand and object motions using T3D (a), VR-GM (b), VR-GMA (c), and VR-Full (d) visualization methods. 
} 
\label{visualization}
\end{figure}

We used traditional and VRMoVi visualizations for hand-object interactions that gradually increased the level of detail being presented:

\begin{itemize}
    \item \textbf{Traditional 3D (T3D).}  Visualization of hand-object interaction using static 3D scatter plots displayed on a 2D screen. An interactive, Python-based data visualization library, Plotly \cite{plotly}, is used, which allows the user to interact with the 3D data using their mouse (zooming or rotating). The motion trajectories of hands and objects are shown in blue and red (Fig.\ref{visualization}(a)). 
    \item \textbf{VRMoVi-GM (VR-GM).}  Visualization of hand-object interaction using animated 3D curves (trajectories) displayed on VR, implemented using the animated general-movement (GM) layer of VRMoVi (Fig.\ref{visualization}(b)). 
    \item  \textbf{VRMoVi-GM+A (VR-GMA).}  Visualization of hand movement and object interaction using animated 3D curves and hand and object models displayed simultaneously on VR. Animated general movement (GM) and hand-object avatar (A) layers of VRMoVi were utilized (Fig.\ref{visualization}(c)). 
    \item \textbf{VRMoVi-GM+A+F (VR-Full).} Visualization of hand movement and object interaction using animated 3D curves, hand and object models, and detailed hand spatial information. We used the full visualization layers available in VRMoVi, including animated general movement (GM), hand-object avatar (A), and fine-grained motion (F) layers. In this method, GM and A layers are displayed simultaneously, followed by the F layer displayed at last to show the exact hand position and rotation at each time step (Fig.\ref{visualization}(d)). 
\end{itemize}

\subsubsection{Hand-Object interaction scenarios}
We selected three hand-object interaction scenarios frequently used in upper limb rehabilitation \cite{ploderer_2016}. These motion data were selected from an open-source VR dataset, OpenNEEDS \cite{emery_2021}, where human activities with object interactions were collected in VR. The hand-object interaction scenarios include two goal-oriented movements, 1) \textbf{Picking-up} an object and 2) \textbf{Tossing} an object, and one open-ended 3) \textbf{Drawing} using an object. Each scenario was represented by each visualization method. 

\vspace{-3em}

\subsection{Data Collection}
\subsubsection{Stage 1}  
For T3D, we built interactive static scatter data plots for each scenario (Fig.\ref{visualization} (a)). For VR-GM and VR-GMA, we created short videos recorded within a VR environment to present a remote monitoring scenario where the patient recorded a video of their progress in VR and sent it to the physical therapist for feedback. In this manner, we can also compare these VR visualizations with T3D on the screen (Fig.\ref{visualization} (b-c)). For each of the visualization methods, to quantify the accuracy of hand-object interaction interpretation across different scenarios, participants conducted the following tasks: \begin{itemize}
    \item \textbf{Task 1 Hand/object:} determine which line represents the hand (as opposed to the object) (RQ1)
    \item \textbf{Task 2 Action:} determine what action is being performed (e.g., writing, catching, holding) (RQ2)
\end{itemize}

After visualizing each condition (i.e., each visualization in each scenario), participants filled out a worksheet to respond to each prompt. Since we gradually added more details each time in the visualization, we did not counterbalance by randomizing the order of using those visualization methods to avoid learnability. Therefore, all participants experienced the visualization methods in the same order, with each condition lasting 10 minutes. 

\vspace{-1em}

\subsubsection{Stage 2}  
To address RQ3, we employed the VR-Full visualization across all scenarios. We created a 20-second video depicting the hierarchical structure of the three layers available in VRMoVi. These layers span from a macro overview of hand-object interaction to the micro details of hand orientation and position at each time step (Fig.\ref{visualization} (b-d)). In the video, the animated general movement (GM) and hand-object avatar (A) layers were concurrently displayed, with the fine-grained motion (F) layer shown subsequently. Following this, in a brief semi-structured interview, participants were asked questions about their comprehension of each VRMoVi layer, difficulties deciphering the motion, and potential enhancements.  

\subsection{Data Analysis}
In Stage 1, we assessed each participant's response as correct, incorrect, or missing. For Task 1, the participant's response was considered correct if they accurately distinguished hand trajectories from object trajectories. In Task 2, we marked a participant's answer as correct if their interpretation of the hand-object interaction aligned with the actual scenario (e.g., "picking-up," "tossing," or "drawing") or was closely related to it. 
Subsequently, we carried out an analysis considering each task and visualization method individually. We first summarized the frequency of correct answers from all non-missing responses. Then, we performed inferential statistics. Given that our data were normally distributed (as per the Shapiro-Wilk test), we employed a repeated measure ANOVA, and a Tukey correction test. p<0.05 was considered to be statistically significant.

During Stage 2, all participant responses were de-identified, documented, and subsequently transcribed. We then implemented methods derived from the thematic analysis \cite{braun_2012}. Participants’ responses were coded and grouped in an affinity diagram to uncover themes. 

\section{Results}

\begin{figure}
\includegraphics[width=\textwidth]{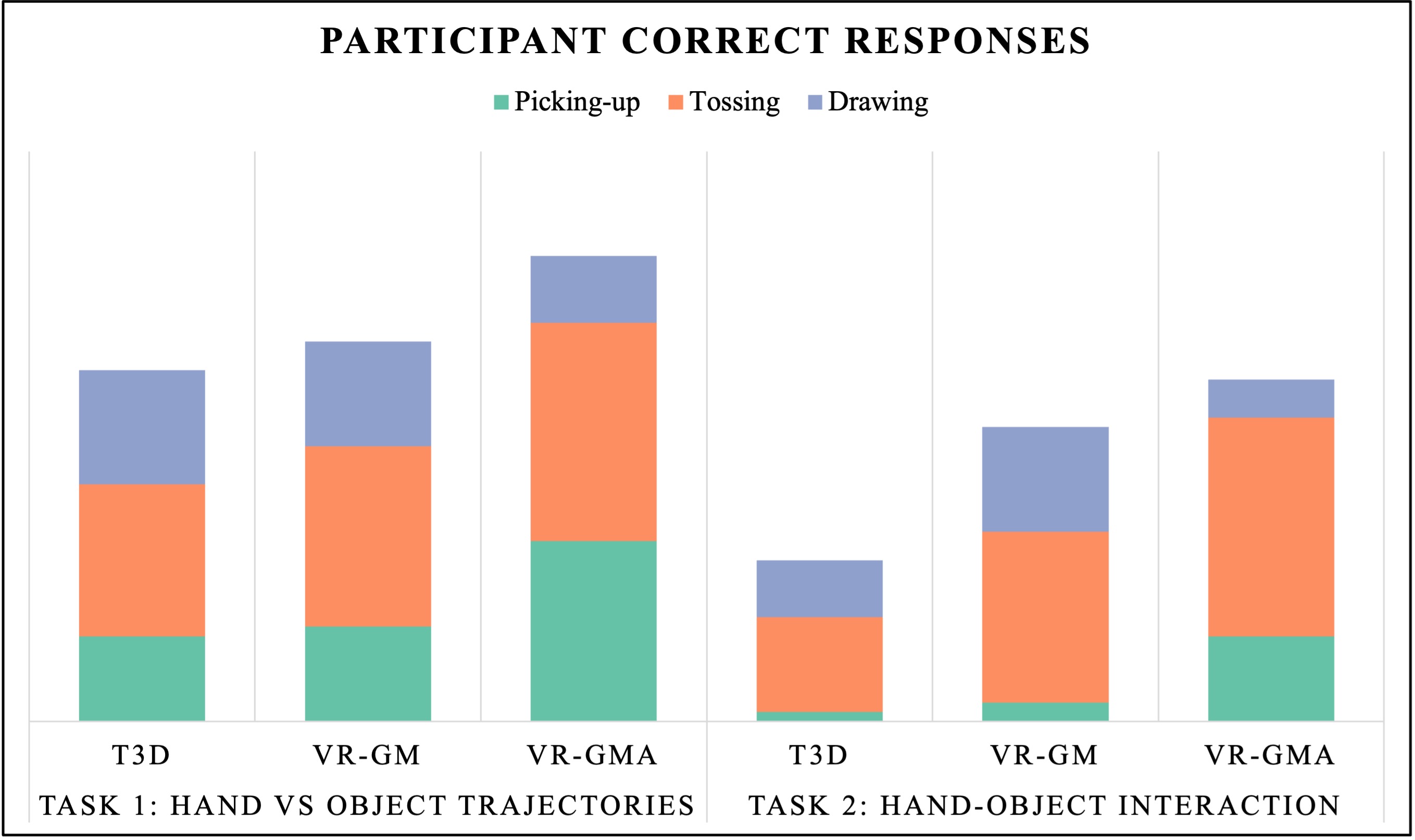}
\caption{The number of correct responses from the participants on Task 1 (left) and Task 2 (right) for each hand-object interaction scenario (picking-up, tossing, and drawing) across visualization methods: T3D, VR-GM, and VR-GMA.} 
\label{graph}
\end{figure}

Fig.\ref{graph} shows the participants’ overall accuracy for Task 1 (left) and Task 2 (right) across all hand-object interaction scenarios. 

\subsection{Task 1: Differentiating Hand and Object Trajectories (RQ1)}
Participants better differentiated the hand from the object trajectories (independently of the action) in the VR+GMA (68\%), than in the VR-GM (56\%) and the T3D (51\%); however, no significant difference was observed (p=0.108).


Specifically, in the picking-up scenario, nearly twice the number of participants were capable of correctly discerning the object’s trajectory from the hand when utilizing VR-GMA (79\%) in contrast to using T3D (38\%; p=0.009) or the VR-GM (42\%; p=0.021). For the tossing scenario, 96\% of participants differentiated the hand and object trajectories using VR-GMA, compared to 67\% with T3D (p=0.029) and 79\% with VR-GM visualization (p=0.298). Conversely, for the drawing scenario, there was no significant disparity among the visualization methods (p=0.310), with almost 50\% of participants able to distinguish the hand and object moving trajectories using T3D, followed closely by VR-GM (46\%), and then VR-GMA (29\%).
 
Overall, the VR-GMA provides better feedback for differentiating hand/object trajectories, especially for task-oriented interaction scenarios, owning to the visual cues it provides (i.e., hand and object avatars), as illustrated in Fig.\ref{graph} (left). 

\subsection{Task 2: Interpreting Hand-Object Interactions (RQ2)}

Regarding participants’ performance in interpreting hand-object interactions, our findings indicated a significant difference across the visualization methods (p=0.022). Nearly twice as many participants accurately identified the hand-object interaction scenarios using VR-GM (44\%) and VR-GMA (54\%) in comparison to T3D (25\%; p=0.031). This suggests that VR-based visualizations provide visual cues about interaction context better than T3D(Fig.\ref{graph}(right)).

Upon analyzing the data for each scenario, significant differences were observed in the tossing (p=0.002), where all participants accurately identified this interaction using VR-GMA (100\%), in contrast to VR-GM (78\%; p=0.348) and T3D (43\%, p=0.001). However, in the drawing and picking-up scenarios, less than half of the participants correctly recognized those interactions. For the drawing scenario, 46\% correctly interpreted it using VR-GM, followed by T3D (26\%) and VR-GMA (17\%). Conversely, in the picking-up scenario, a higher proportion (43\%) correctly identified it using VR-GMA, while only 2 and 1 participants managed to identify this interaction using VR-GM and T3D.

In summary, owning to the animated trajectory and hand-object interaction incorporated into the VR-based visualization methods with VR-GMA superior to VR-GM, participants were more successful in recognizing the hand-object interaction scenarios than those utilizing T3D.

\subsection{Challenges and Opportunities for VR Visualization for Hand-Object Interaction (RQ3)}
Our findings in Stage 2 suggest that VR visualizations of hand-object interaction effectively convey information about representations, particularly when the animation, visual cues, and level of detail are carefully crafted. 
\vspace{-1em}
\subsubsection{Translating Visualization into Meaningful Information.} 

Participants identified certain motion characteristics they considered important for the visual representation of hand movement. Some of these features, such as the trajectory and angles of movements, can be effectively depicted in traditional 3D and VR visualizations, as they said: "\textit{[Visualization can help] to understand the path hands take for various actions.}"  

As expected, participants highlighted features that VR animations could represent effectively, including velocity, direction, and depth: "\textit{You could see how speed and positioning change the visuals.}" Particularly for hand direction, participants discussed the potential for showing the orientation of a hand, object, or both synchronized: "\textit{What direction the hand is facing,}" and augmented with VR capabilities "\textit{Show hand movement and “ball” movement then both at the same time.}" 

Although VR-Full visualization uses arrows overlaying the trajectory path to indicate hand rotations during movement, participants associate them with features like force, energy, or velocity. Complementing the animation of arrows, the motion of the dots was linked to frequency or velocity: "\textit{Dots that are slower represent slower movement...}"

Overall, features such as direction, velocity, and movement synchronization should leverage the unique capabilities of VR environments to enhance visualization. This approach could enable clinicians to offer remote advice to patients more effectively.
\vspace{-1em}
\subsubsection{Improvements for Representation.}  Participants identified that the contrast in the existing visualization might not be accessible: "\textit{Environment is really dark, hard to see stuff, could be in a lighter setting.}" Therefore, suggestions included adopting a lighter setting or offering options to alter the colors. Additionally, to enhance accessibility further, participants recommended: "\textit{Provide more context to what the action is by using different objects and colors. Add a timer so you can see how much time elapsed and how quickly the action is occurring.}"

On the other hand, animation was perceived as a beneficial and simplified understanding of the movement: "\textit{The animation made the motions easier to understand rather than a still diagram.}" Participants concurred that real-time animation could better delineate the movement's start and end points, the intended action to be performed, and the movement's speed. However, they suggested these detail levels could be augmented by: "\textit{Add additional views for motion, size, and angle that can be interpreted for what the user is achieving with or without the object.}" Furthermore, explicit symbols could enhance trajectory representation, such as hand and object symbols: "\textit{Add a recording of VR hand and real-life hand,}" as well as clearly defined start and end points, ensuring a one-to-one correspondence of movements. 

Based on participant feedback, aspects such as accessibility, animation, visual cues, and detail levels should be refined to provide clinicians with an accurate representation of the movements. 



\section{Discussion}
VR visualizations have shown great promise in offering insightful data about movements \cite{kloiber_2020,bschel_2021,reipschlger_2022}, holding the potential to enhance the process of remote monitoring in physical therapy by supplying clinicians with detailed information about patient movements, thereby facilitating remote feedback. 

Our study demonstrated that integrating animation and 3D  symbols significantly improved the understanding of movement actions and differentiated hand and object trajectories, particularly for goal-oriented movements like tossing or picking up an object. 
Therefore, delivering intelligible and precise visualizations of these movements is crucial. However, adding details might confuse open-ended movements, such as drawing, suggesting a simpler animation approach. An alternative strategy might entail incorporating multiple perspectives, thus offering a "zoom-in" function to display intricate motion details for open-ended or complex hand actions such as "drawing." This would allow the user to examine a specific portion of the data closely.

Our qualitative analysis showed that enhancing hand-object interaction trajectories with detailed visual symbols facilitates the translation of movement data into meaningful information. Therefore, using symbols \cite{rado_presentation_2009,moya_2011} delivers visual cues about the position, direction, angle, speed, and depth, enabling physical therapists to make informed decisions when providing remote patient care. When designing these visualizations, it is crucial to consider factors such as the contrast between the background and 3D visualizations to ensure clear differentiation. Moreover, offering explicit labels and allowing user customization of displayed information can further improve motion understanding.

Despite achieving our objectives, we acknowledge some inherent limitations. Our findings are based on a 2-stage study conducted with participants from visualization-experienced backgrounds who were asked to conduct tasks related to specific movements. Therefore, to generalize the outcomes for the physical therapy field, in future work, we plan to engage physical therapists to gain feedback, refine our current visualization design, and assess the visualization's effectiveness in delivering meaningful information via qualitative and quantitative studies. Furthermore, we plan to increase hand-object motion scenarios and incorporate technical metrics such as accuracy and system performance.  

Overall, considering our overarching goal of providing effective 3D visualizations for assessing object manipulation skills in remote physical therapy, we believe our research results signify an invaluable preliminary step in demonstrating the potential of VR visualization in supporting remote physical therapy. 

\section{Conclusion}
This paper presents the development of VRMoVi, a VR-based visualization system that incorporates multiple levels of detailed 3D visualization features to depict human motion. Our results show that VR visualizations can potentially improve the understanding of hand-object interactions and enhance the interpretation and analysis of physical therapy remotely. The future work will evaluate the effectiveness of the proposed VRMoVi visualizations for assessing object manipulation skills in remote physical therapy, working with physical therapists. 

%
%

\subsubsection{Acknowledgements}
This research was funded by the 2022 Research Seed Fund of Dr. Trudi Qi, the 2020-22 Research Startup Fund of Dr. Franceli Cibrian, and the Robert A. Day Undergraduate Research Grant of Meghna Raswan at Chapman University.

%
%
%
 \bibliographystyle{splncs04}
 \bibliography{references}

\end{document}